# Size effect in ion transport through angstrom-scale slits

A. Esfandiar,[1] B. Radha,[1,2] F. C. Wang,[1,3] Q. Yang,[2,4] S. Hu,[2] S. Garaj,[5,6,7] R. R. Nair,[2,8] A. K. Geim,[1,2] K. Gopinadhan[1]

[1]School of Physics & Astronomy, University of Manchester, Manchester M13 9PL, UK
[2]National Graphene Institute, University of Manchester, Manchester M13 9PL, UK
[3]Chinese Academy of Sciences Key Laboratory of Mechanical Behavior & Design of Materials, Department of Modern Mechanics, University of Science & Technology of China, Hefei, Anhui 230027, China
[4]Key Laboratory of Advanced Technologies of Materials, School of Materials Science & Engineering, Southwest Jiaotong University, Chengdu, Sichuan 610031, China
[5]Department of Physics, National University of Singapore, 117542 Singapore
[6]Centre for Advanced 2D Materials, National University of Singapore, 117546 Singapore
[7]Department of Biomedical Engineering, National University of Singapore, 117575 Singapore
[8]School of Chemical Engineering & Analytical Science, University of Manchester, Manchester M13 9PL, UK

*It has been an ultimate but seemingly distant goal of nanofluidics to controllably fabricate capillaries with dimensions approaching the size of small ions and water molecules. We report ion transport through ultimately narrow slits that are fabricated by effectively removing a single atomic plane from a bulk crystal. The atomically flat Å-scale slits exhibit little surface charge, allowing elucidation of the role of steric effects. We find that ions with hydrated diameters larger than the slit size can still permeate through, albeit with reduced mobility. The confinement also leads to a notable asymmetry between anions and cations of the same diameter. Our results provide a platform for studying effects of Å-scale confinement, which is important for development of nanofluidics, molecular separation and other nanoscale technologies.*

Many natural materials and phenomena involve pores of angstrom dimensions[1-4]. Examples are Å-size ion channels in cellular membranes, which are crucial for life's essential functions, and ion-exchange membranes used in desalination, dialysis and other technologies[1-4]. To mimic and better understand functioning of such ion-transport systems, it is desirable to controllably fabricate and investigate artificial channels with similar dimensions. Unfortunately, channels fabricated using standard lithography techniques and conventional materials are limited in size by intrinsic roughness of materials' surfaces, which typically exceeds the hydrated diameter $D_H$ of small ions by at least an order of magnitude[5,6]. Ion transporters with nanometer dimensions have also been demonstrated, including smooth-walled carbon and boron-nitride nanotubes[7-9], nanopores made in monolayers of MoS$_2$ and graphene[10-15], and interlayer passages in graphene oxide laminates[16-18]. These nanochannels still have sizes considerably exceeding those typical for inorganic ions and suffer from the presence of many defects and, especially, built-in electric charges[5,13,19,20]. Although the latter systems delivered many insights, it has often been difficult to disentangle various mechanisms that contribute to ion transport through them, including exit-entry effects, surface charges, steric exclusion and others. Recently, we reported atomically-flat slits down to several Å in height, which were controllably made by van der Waals (vdW) assembly[21]. Unlike quasi-one-dimensional nanotubes and biological channels, our capillaries are two-dimensional and, in contrast to synthetic and biological ion transporters, have chemically inert and atomically smooth walls. In this paper, we investigate ion transport under such ultra-strong confinement.

The slit devices were fabricated following the recipe described in refs. 21 & 22. In brief, our channels comprised two relatively thick (∼ 100 nm) crystals obtained by mechanical exfoliation. For the present study, we chose graphite, hexagonal boron nitride (hBN) and molybdenum disulfide ($MoS_2$). The crystals were placed on top of each other, separated by stripes of bilayer (2L) graphene or monolayer (1L) $MoS_2$ which served as spacers (fig. S1). The assembly was kept together by vdW forces, and the resulting channels had the height $h$ of ∼ 6.6 and 6.7 Å, given by the vdW thicknesses of 1L $MoS_2$ and 2L graphene, respectively (Fig. 1a). This height is comparable to, e.g., the diameter of aquaporins[1] and is our smallest achievable $h$ because slits with thinner spacers were intrinsically unstable, collapsing due to vdW attraction between opposite walls[21]. The reported Å-slits had the width $w \approx 0.13$ μm and the length $L$ of several μm. The tri-crystal stack was placed on top of a silicon nitride membrane with a rectangular opening of $3\times25$ μm$^2$, which served as a mechanical support and a partition between two liquid reservoirs (fig. S2). The reservoirs were thoroughly isolated from each other to ensure that ion transport occurred only through Å-slits (Fig. 1a).

First, we tested individual slits and measured their ionic conductivity using KCl solutions with the same molar concentration $C$ in both reservoirs (fig. S3). The recorded current-voltage (*I-V*) characteristics were linear at small biases (< 30 mV) exhibiting only slight nonlinearity over the studied voltage range up to ± 0.2 V. The linear-response conductance $G$ was ∼ 0.5 nS for $C$ = 1 M, in agreement with the known bulk conductivity of the KCl solution for the given geometry. By decreasing the KCl concentration, we rapidly reached our detection limit set by leakage currents (fig. S3). The typical leakage corresponded to ∼ 10 pS, as found using reference devices without spacers. To increase sensitivity, we therefore opted to work with devices containing 200 slits in parallel[22]. At large $C$, they exhibited the same conductance per channel as that found for single-channel devices (fig. S3). Typical *I-V* characteristics for 200-channel devices are shown in Fig. 1b and fig. S3. Their linear-response $G$ is plotted in Fig. 1c. At high salt concentrations, $G$ is proportional to $C$ and agrees with the KCl bulk conductivity inside the Å-slits. $G$ starts to deviate from the linear dependence at ∼ $10^{-2}$ M and then saturates to the leakage level of ∼ 10 pS (Fig. 1c). As an additional test, we studied large ion exclusion using tetra- methyl, ethyl and butyl ammonium chloride solutions. No ionic current could be detected above the leakage limit for salts with $D_H$ > 13 Å (fig. S4). We carried out similar experiments for Å-slits with hBN and $MoS_2$ walls (Fig. 1d) and found $G(C)$ similar to that for graphite walls. However, the saturation at low $C$ occurred to notably higher values of $G$. Such behavior is well known in nanofluidics and attributed to electric charges present on capillary surfaces[5,6,8,15]. We evaluated the surface charge density $\Theta$ using the standard analysis[22], which yielded ∼ 120 and 300 μC m$^{-2}$ for hBN and $MoS_2$ surfaces, respectively. For graphite walls, the finite leakage allowed only the upper-bound estimate $\Theta \leq 20$ μC m$^{-2}$. These values are 3-4 orders of magnitude smaller than those reported for carbon and hBN nanotubes (∼ 10 and 100 mC m$^{-2}$, respectively[8,9]), silica channels, graphene oxide laminates and pores in monolayer crystals, which typically exhibit $\Theta$ ∼ 100 mC m$^{-2}$ (refs. 5, 13, 15 & 17). Even for our Å-slits with $MoS_2$ walls, an average distance between charged defects is > 20 nm. We also find no difference in $\Theta$ between channels having side walls (spacers) made from graphene and $MoS_2$ (Fig. 1c) as expected for the low aspect ratios $h/w$ < 0.01. We believe that the low surface charge density in our capillaries is due to their extreme cleanliness. Their top and bottom walls contain virtually no surface defects being mechanically exfoliated from quality bulk crystals[21,22], in contrast to low-dimensional materials grown by, e.g., chemical vapor deposition.

The employed surface-charge model provides good description of the experiment at both high and low $C$ (dashed curve in Fig. 1c; fig. S5) but the measured $G$ is notably higher than expected for intermediate $C \approx 10^{-2}$ to $10^{-4}$ M. Such deviations are usually assigned to the surface charge that depends on $C$ (so called charge regulation model[9]). The solid curves in Fig. 1c, d show that the model describes our results better. The variable charge can be attributed to OH⁻ adsorption on capillary walls[8,9]. This agrees with our measurements of KCl conductance under different pH from 2 to 10, which showed[22] that $G$ was practically constant under acidic conditions but rapidly increased for high basic pH (fig. S6). Note that both constant and variable charge models yield the same intrinsic charge density.

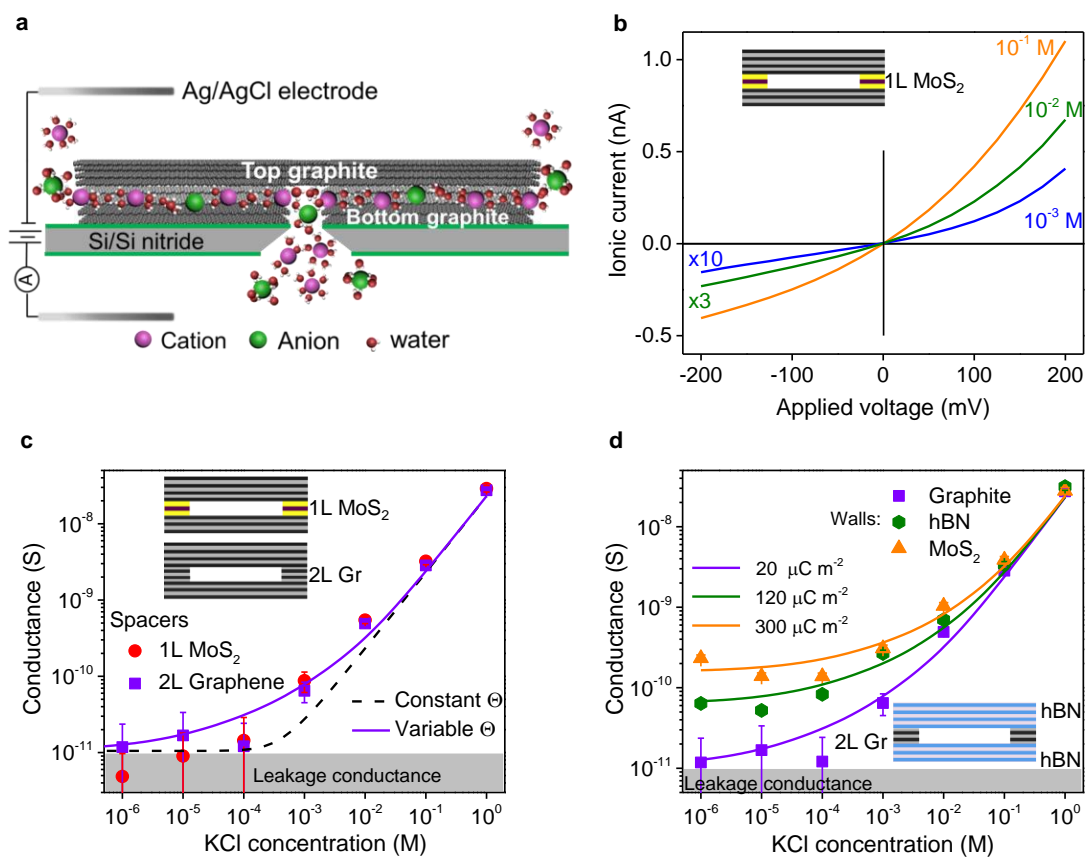

**Figure 1| Ion transport under Å-scale confinement. a,** Schematic of our measurement setup. **b**, $I$-$V$ characteristics of a device with 200 channels in parallel; $w \approx 0.13$ μm, $L \approx 7$ μm, 1L MoS$_2$ spacers. KCl concentrations vary from $10^{-3}$ to $10^{-1}$ M. For clarity, the curves for low $C$ are magnified by the color-coded factors. **c,** Conductance for two representative devices with 2L graphene and 1L MoS$_2$ as spacers (symbols). For KCl concentrations $\leq 10^{-4}$ M, the measured $G$ was comparable to typical electrical leakage, as indicated in grey. **d,** Conductance of slit devices made from graphite, hBN and MoS$_2$, using 2L graphene spacers; $L \approx 7$ μm. In (c) the dashed curve is a fit assuming a constant surface charge, whereas the solid curves in (c) and (d) represent fits using the variable charge model. Error bars: Accuracy of determining $G$ for individual $I$-$V$ curves. Insets: Schematics of the used slits.

The small Θ and little dependence on walls' chemistry provide a unique opportunity to examine more subtle effects in ion transport. Because $h$ is comparable to $D_H$ for small ions, we investigated whether, in addition to the found complete exclusion of large (> 13 Å) ions, our Å-slits provide any size effect for common inorganic salts, as widely discussed in literature and important for applications[1-3]. To this end, several chloride solutions were chosen with cations' $D_H$ ranging from ~ 6.6 to 12.5 Å. Chloride's $D_H$ is 6.6 Å. Despite reaching the limit $h < D_H$, our channels exhibited no abrupt steric exclusion (Fig. 2a; fig. S4b), contrary to what is often assumed when modeling ion transport. As reported above (Fig. 1), conductivity σ of KCl, where both ions had $D_H \approx h$, changed relatively little with respect to its bulk conductivity. The chloride solutions with larger cations exhibited a notable reduction in their σ which reached a factor of 4 for $Al^{3+}$ and ~ 50 for tetramethylammonium (their $D_H$ are ~ 1.5 and 2 times larger than $h$, respectively). These observations clearly show that ions under confinement do not act as hard balls but are able to partially shred or flatten their hydration shells[14,23].

It is known that edges of nanopores and nanotubes have a profound effect on their ionic conductance[10,11,13,15,18]. To find out whether similar entry-exit effects contribute to ion transport through our Å-slits, we studied dependence of their $G$ on the channel length $L$. An example is shown in Fig. 2b for two chloride solutions. The measured resistance $1/G$ increases linearly with $L$ and, within our accuracy, the linear fits extrapolate to zero. This indicates little contribution from entry-exit barriers, proving that the conductance is dominated by ion diffusion inside the slits.

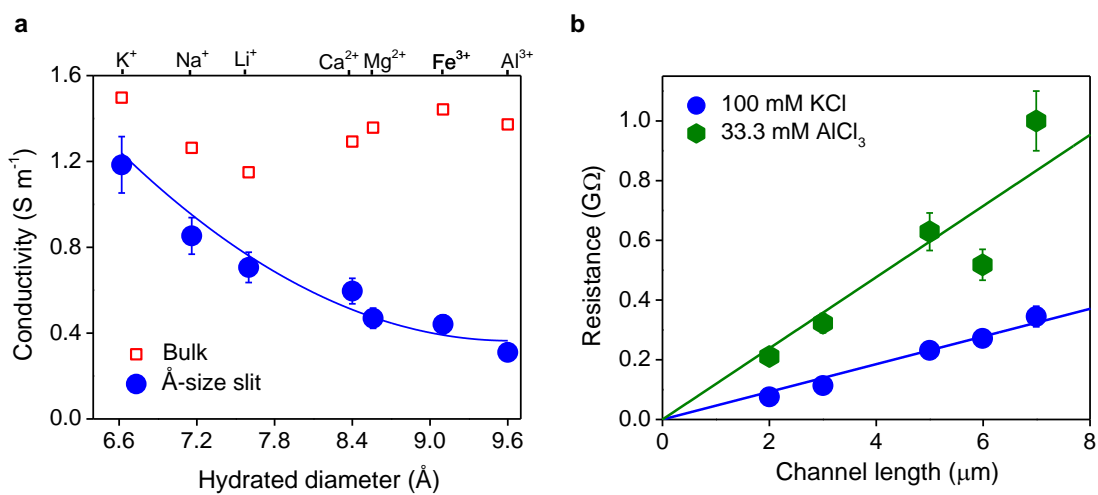

**Figure 2| Size effect in ionic conductivity. a,** Conductivity of various 0.1 M solutions for a device with graphite walls and 1L $MoS_2$ spacers (blue circles). Chlorides' cations are listed along the top axis, and their $D_H$ are given against the bottom axis. Solid curve: Guide to the eye. Open squares: Salts' bulk conductivity. **b,** Å-slits' resistance $1/G$ as a function of $L$ (graphite walls; 2L graphene spacers). Solid lines: Linear fits. Error bars: Standard deviations in determining $G$ from $I$-$V$ curves.

To gain more information about influence of Å-scale confinement on ion transport, we performed drift-diffusion experiments[8,13,15,17]. The two reservoirs were again filled with various chloride solutions but now in different concentrations. Specifically, we used 10 and 100 mM solutions in the permeate and feed reservoirs, respectively (Fig. 3a). Because cations and anions generally diffuse at different rates, a finite ion current arises even in the absence of applied voltage and, consequently, $I$-$V$ curves become shifted along the voltage axis (Fig. 3a). A positive current at zero $V$ corresponds to higher mobility of anions, $\mu^-$, compared to that of cations, $\mu^+$. For example, the curves in Fig. 3a show that $K^+$ and $Al^{3+}$ diffuse through our Å-slits faster and slower than $Cl^-$, respectively. The zero-current potential $E_m$ allows us to find the mobility ratio, $\mu^+/\mu^-$, (refs. 13 & 17) using the Henderson equation[25]

$$\mu^+/\mu^- = -\frac{z_+}{z_-}\frac{\ln(\Delta) - z_- FE_m/RT}{\ln(\Delta) - z_+ FE_m/RT} \qquad (1)$$

where $z_+$ and $z_-$ are the valences of cations and anions, respectively, $F$ is the Faraday constant, $R$ the universal gas constant, $T$ = 300 K, and $\Delta$ is the ratio of $C$ in the feed and permeate containers. In our experiments, $\Delta$ = 10 and $z_-$ = -1. Fig. 3b plots $\mu^+/\mu^-$ obtained using eq. (1). The mobility ratio changes by an order of magnitude with increasing $D_H$ from $K^+$ to $Al^{3+}$ but is remarkably indifferent to the wall material. We also used reference capillaries with the size $\gg D_H$ where no steric effects were expected[22]. The latter devices exhibited $\mu^+/\mu^-$ values very close to those reported in the literature for bulk solutions, confirming accuracy of our analysis for Å-slits (figs. S7-S9).

It is more informative to find $\mu^+$ and $\mu^-$ rather than their ratios. To this end, we measured conductivity of various chloride solutions (as in Fig. 2a) using relatively high $C$ = 0.1 M so that the surface-charge contribution could be neglected. The conductivity can then be described as $\sigma \approx F(c_+\mu^+ + c_-\mu^-)$ where $c_\pm$ are the concentration of anions and cations. Combining the latter equation with the found $\mu^+/\mu^-$, we obtained $\mu^+$ and $\mu^-$. Their values are plotted in Fig. 3c. The mobility of $Cl^-$ varies little for different salts (within $\pm 15\%$) but its absolute value under the confinement becomes $\sim$ 3 times smaller than in bulk solutions (Fig. 3c). In stark contrast, the cations exhibit a decrease in mobility by a factor of $\sim$ 10 with increasing $D_H$ from $K^+$ to $Al^{3+}$. Interestingly, despite $K^+$ and $Cl^-$ exhibit the same $D_H$ and similar mobilities in bulk solutions[22,24], the mobility of $K^+$ remains practically unaffected by the confinement whereas $Cl^-$ becomes 3 times less mobile (Fig. 3c).

While the reported exclusion of very large ions from our Å-size slits is generally expected, it is rather surprising that small ions exhibit only the modest suppression of their mobility if the sieve size $h$ becomes notably smaller than $D_H$ (c.f. refs. 16-18). A number of molecular dynamic simulations previously suggested that under such 'quantum' confinement, ions can reconfigure their hydration shells that become effectively squashed[14,23]. A qualitative measure of the difficulty for such water-molecule rearrangements around ions is their hydration energy that describes the energy accumulated in hydration shells and increases with $D_H$ (refs. 26 & 27). We expect that coefficients describing ions' kinetics depend exponentially on the activation energy barrier presented by involved dehydration processes. This is in conceptual agreement with the fact that the measured $\mu^+$ evolve approximately exponentially with the hydration energy as indicated by the use of two y-axes in Fig. 3b. Furthermore, the suppression of chloride's mobility with respect to its bulk value is puzzling, particularly because $K^+$ ions, having the same $D_H$, exhibit no discernable change. We attribute the asymmetry to different polarization of water molecules around cations and anions[22]. The hydration shell of $K^+$ has hydrogen atoms pointing preferentially outside (fig. S10). In contrast, the exterior of $Cl^-$ is covered with $OH^-$ groups. It is known[28,29] that both graphene and hBN polarize water molecules so that $OH^-$ groups are directed preferentially towards the surfaces (also agrees with the pH dependence reported in fig. S6). This suggests that anions should exhibit stronger interaction with graphene and hBN walls, compared to cations of the same size[28,30]. This would result in extra friction of $Cl^-$ against the walls and, consequently, reduce its mobility, in agreement with the observed cation-anion asymmetry. Further theory analysis is required to describe the reported dehydration and asymmetry effects quantitatively.

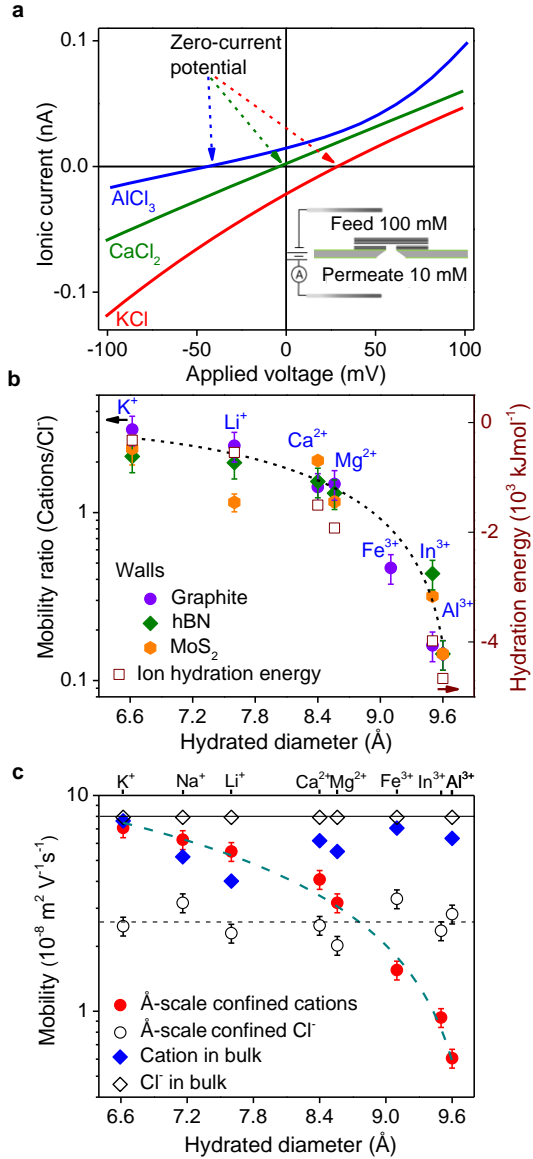

**Figure 3| Ion mobility under Å-scale confinement. a,** Examples of *I-V* characteristics for various chloride solutions under the concentration gradient Δ = 10 (device with graphite walls and 1L MoS$_2$ spacers). Inset: Schematic of the drift-diffusion measurements. **b,** Mobility ratio $\mu^+/\mu^-$ as a function of cation's $D_H$ for slits made from graphite, hBN and MoS$_2$ (color coded) using 2L graphene spacers. The ions' hydration energy is shown by the empty squares against the right axis. **c,** Ion mobility under the confinement as a function of $D_H$ (circles). Shown is the most complete data set (graphite walls). Other walls yielded similar values. Diamonds: Literature values[24] for $\mu^+$ and $\mu^-$ in bulk solutions. Curves in (b) and (c): Guides to the eye.

To conclude, our atomically flat Å-size slits exhibit, in the first approximation, little chemical interaction with ions and act purely as a geometric confinement. The observed changes in small ions' mobility can be explained by distortions of their hydration shells, which become increasingly costly in terms of energy with increasing the ion diameter. Our results imply that any feasible confinement is unlikely to provide high selectivity between small ions, and living and artificial systems have to rely on strategically placed electric charges inside channels or at their entries.

**SUPPLEMENTARY INFORMATION**

**S1 Fabrication procedures**

Our Å-slit devices consisted of three crystals prepared by mechanical exfoliation of bulk layered materials (fig. S1a). First, a crystal of graphite, hBN or $MoS_2$ (~ 10 - 15 nm in thickness) was transferred onto a silicon nitride membrane with a rectangular hole of about 3×25 µm² in size, which was prepared in advance using a standard Si wafer covered with silicon nitride[21]. This crystal is referred to as a bottom layer. It is dry-etched from the backside using the hole in the silicon nitride membrane as a mask. Next, either bilayer graphene or monolayer $MoS_2$ was exfoliated onto an oxidized Si wafer (300 nm of $SiO_2$) and patterned using electron beam lithography and dry etching, which resulted in an array of stripes such those as shown in fig. S1b. The stripes were separated by ~130 nm and served as spacers. The spacer array was subsequently transferred onto the bottom layer, aligning the stripes perpendicular to the long-axis of the rectangular hole. The atomic force microscopy (AFM) images revealed a thickness of ~ 7 Å for both 2L graphene and 1L $MoS_2$ after the spacers were placed on the bottom graphite layer, as expected from their vdW thickness (inset of fig. S1b). The resulting stack was again dry-etched from the backside to remove the spacer stripes from inside the hole (fig. S1a). Then, another thin crystal of graphite, hBN or $MoS_2$ (~ 70 - 100 nm thick) was transferred on top of the stack. This crystal (referred to as top) completed the Å-size slit, providing its enclosure from the top. The layer's thickness provided enough mechanical rigidity to avoid sagging of the slits in air[21] or their swelling in water.

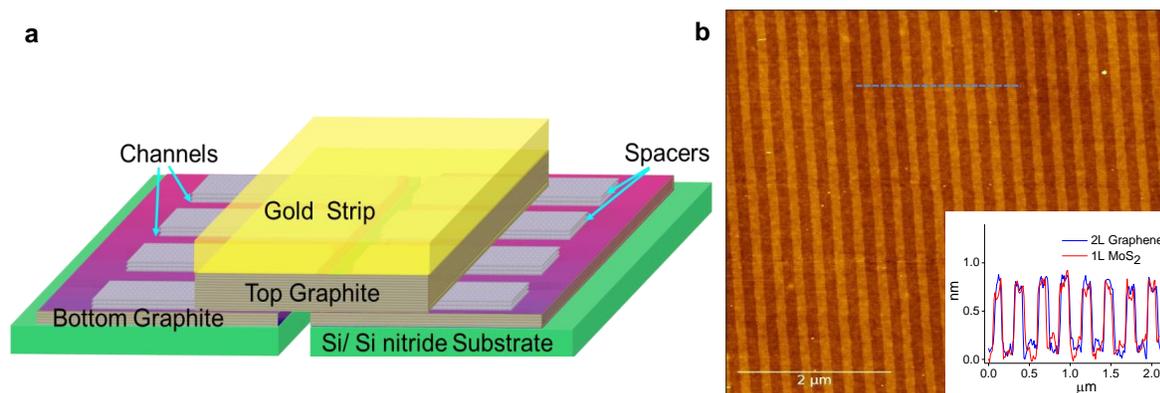

**Figure S1| Basic design of our Å-slit devices. a,** Their schematic. **b,** AFM images of bilayer graphene spacers on top of the bottom graphite layer. Inset: Height profiles yield an estimate of ~ 7 Å for the thickness of spacers made from both 2L graphene and 1L $MoS_2$ (the blue line in (b) shows the scan position for the corresponding trace in the inset).

After each transfer, we annealed our graphite and hBN devices in an $Ar/H_2$ atmosphere at 400°C for 3 hours to remove any possible contamination and polymer residue. For devices with walls or spacers made from $MoS_2$, a lower annealing temperature of 300°C was employed to avoid possible degradation of the latter material. We found that the multiple annealing steps were crucial for obtaining good functional devices. Obviously, the annealing helped to reduce the number of adsorbates. However, we believe that there is also another cleansing mechanism involved. Indeed, all surfaces exposed to air are normally covered with adsorbed hydrocarbons that can also provide charge traps. Our Å-size channels are too narrow to allow typical polymer molecules inside. Those are squeezed away which results in self-cleansing, similar to the mechanism reported in ref. 31. No such self-cleansing is expected for nanotubes of a few nanometers in diameter because a typical diameter of adsorbed hydrocarbon molecules used in microfabrication technologies is ~ 2 nm.

The fabrication procedures described above were introduced in our previous work[21]. Since then further improvements in the design have been made, aiming to improve the success rate and reliability of the technology. In particular, we addressed the problem that, although top-layer crystals often had sharp edges, in some cases their thickness decreased only gradually, step-like. As reported previously[21], crystals thinner than 10 – 20 nm can deform and sag into capillaries blocking them. In this work, to avoid such blockage of channel entries by thin edges, a gold strip (5/150 nm of Cr/Au) was placed to cover the top crystal and aligned along the opening in the silicon nitride wafer (fig. S1a). Typically, the strip had a width of ∼ 20 μm and was made by photolithography. It served as a mask for dry etching that lasted long enough to open fresh entries to the channels (fig. S2). This fabrication step notably improved the reproducibility of our measurements. It also provided the same channel length $L$ for all 200 slits in parallel, which allowed us to avoid averaging in defining an effective length of our devices[21]. Moreover, the gold strip provided additional clamping of tri-crystal vdW stacks to the silicon nitride membrane (fig. S2c-d), which stopped them from being lifted during emersion in a liquid or electrical measurements. Such lifting occasionally happened with our previous devices and high concentration gradients ∆ (see section S6 below).

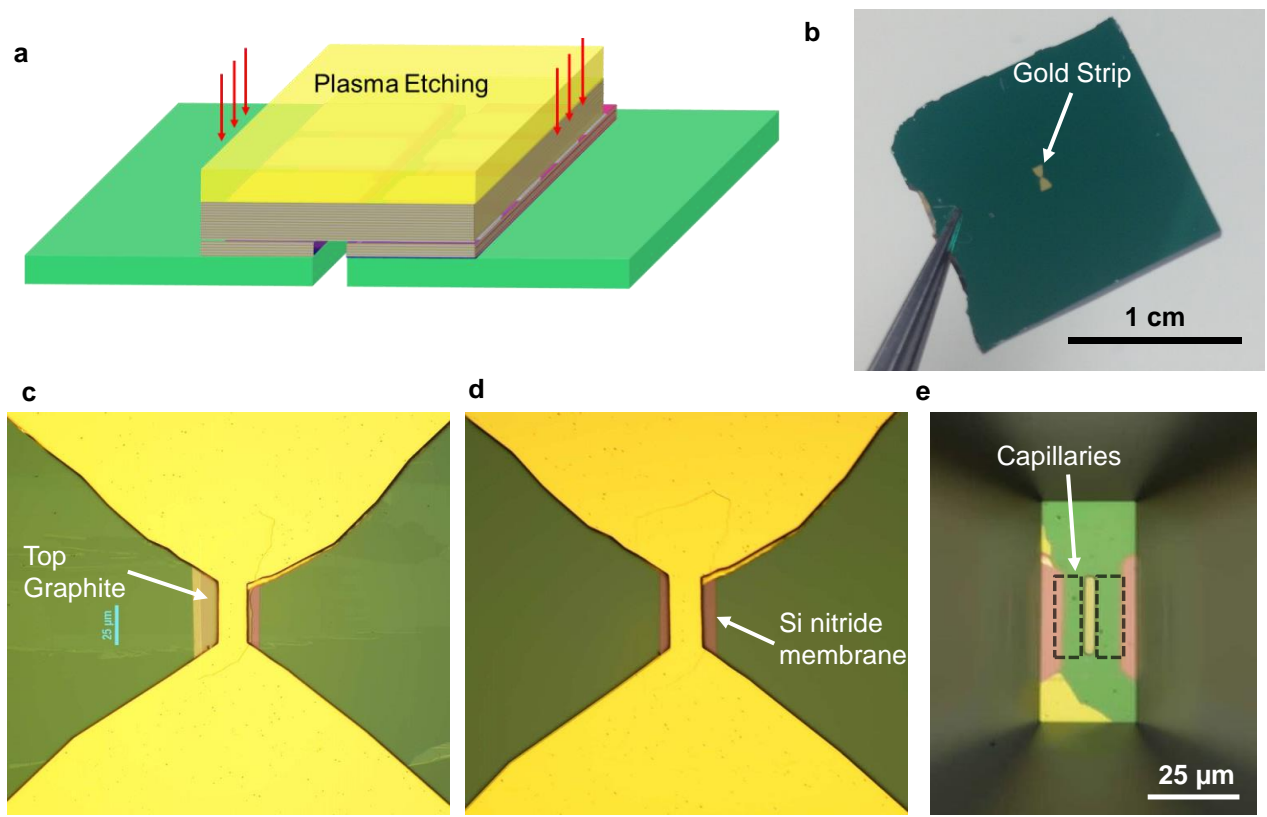

**Figure S2| Improving design and reliability. a,** Schematic of using a gold strip to open or freshen up slits' entries. **b,** Overview image of the final device assembled on a Si wafer. (**c** and **d**) Optical micrographs of our devices with the top Au strip before and after dry etching, respectively. **e,** Optical image from the backside, through the hole etched in the Si wafer. Here the clean part of the silicon nitride membrane is seen in pink. It is partially covered by the bottom graphite, which results in the green color. The top gold layer not covering the tri-layer stack is seen in yellow. The position of Å-slits is indicated by the dashed rectangles, each containing 100 channels.

**S2 Ion transport measurements**

Our electrochemical cell was custom-made. It was machined from PEEK (polyether ether ketone) and consisted of two reservoirs with a capacity of 5 ml each. The Si wafer containing an Å-slit device (as shown in fig. S2b) was placed between the reservoirs and sealed using acid-resistant O-rings. The relatively large volume of the reservoirs provided an easy escape for air bubbles formed during solution filling. Remaining bubbles, if any, were removed by repetitive rinsing of the Si wafer from both sides. To this end, we sequentially filled the reservoirs with pure ethanol, ethanol in deionized (DI) water (1:1) and, finally, DI water only. The first two steps were used to improve the wetting of surfaces. The measured pH of the DI water (*Millipore*) was ~6. The deviation from the ideal pH of 7 is attributed to $CO_2$ dissolution from air, which is a well-known effect. We also tested water from other sources and the results were same. For ion transport measurements, the reservoirs were filled with chloride solutions in chosen concentrations, which substituted the DI water.

*I-V* characteristics were measured by *Keithley* 2636B SourceMeter and recorded using *LabVIEW*. Normally, Ag/AgCl electrodes were employed but, in the case of different salt concentrations, to eliminate the potentials arising from redox reactions on the electrodes, we used standard electrodes (saturated Ag/AgCl salt bridge electrodes from *HANA Instruments*). At low salt concentrations (< $10^{-4}$ M), we used long acquisition times of up to 300 s to avoid hysteresis in *I-V* characteristics. The maximum applied voltage was limited to be ± 0.2 V because higher biases resulted in hysteresis, irreproducible changes and eventual destruction (delamination) of our devices. During measurements of the reported concentration dependences, the sequence of used solutions was always from low to high *C*. Once a measurement series with different *C* was completed, the cell was thoroughly washed with DI water to remove any residual salts. This procedure was repeated until the device showed the conductance characteristic for pure DI water. After the cleaning, a device could be used for measurements with other salt solutions.

We normally stored our devices in water to avoid possible contamination. After each series of measurements, we checked the devices' integrity and general reproducibility by measuring their conductance inside a 1M KCl solution. We found that the conductance was always the same, unless a device was accidentally delaminated or contaminated. The constant KCl conductance before and after the measurements indicates that slits did not gradually swell during their investigation.

The usual life time of devices stored in water was a few months but slits with hBN walls survived longer than the others. To increase the lifetime and stability even further, we lately developed additional clamping using polydimethyl siloxane (PDMS). To this end, a 20 μm thick PDMS film with a circular opening of 50 μm in diameter was mechanically transferred on top of the device assembly, exposing only its functional area. Devices clamped with PDMS survived for more than a year without showing any degradation in their conductance.

In total, we fabricated more than 40 devices with 2L graphene and 1L $MoS_2$ spacers, and approximately 60% of them were fully operational. Let us also mention that, for thicker spacers (not reported in this paper), we have achieved a higher success rate up to 90%.

Each salt solution was tested using at least three different devices, and sample-to-sample variations are reflected in error bars shown on the plots.

**S3 Single-channel conductance**

Typical *I-V* characteristics for devices with 1 and 200 channels are shown in figs. S3a and S3b, respectively. Their linear-response conductances calculated per channel are compared in fig. S3c. At high *C*, both devices exhibited the same per-channel conductance, as expected for slits exhibiting same characteristics. This proves the correct scaling of the measured conductance with the number of Å-slits involved. However, our one-channel devices reached the limit given by electrical leakage already at relatively high *C* of about 0.1 M. For 200-channel devices, the saturation level per channel was consequently two orders of magnitude lower, which allowed us to extend the studies to lower concentrations.

In the case of one-channel devices, we found it helpful to create extra nano-cavities around the main channel. To this end, we fabricated an array of spacers such as shown in the inset of fig. S3a. The resulting cavities were disconnected from the conducting channel and served to collect occasional hydrocarbon and other contamination between top and bottom crystals. This design improved our success rate of getting working devices. For one-channel devices, we also used short channels ($L \approx 2$ μm) to increase the ionic current as much as possible.

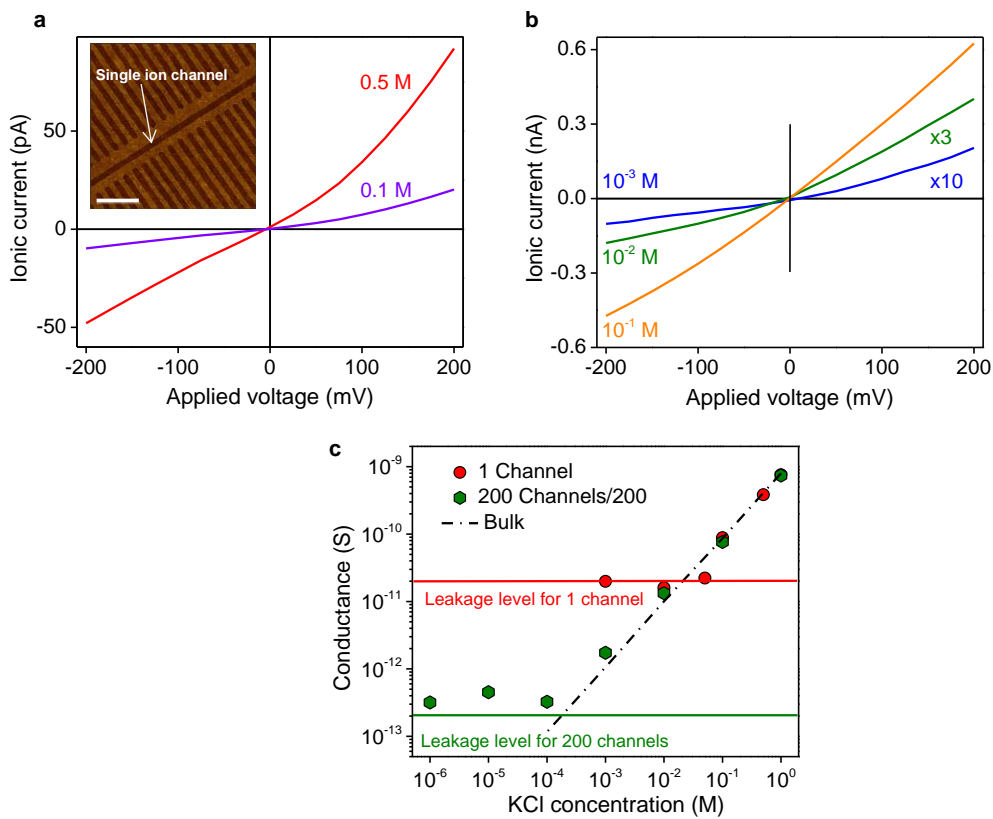

**Figure S3| Scaling of ionic conductance with the number of channels. a,** *I-V* characteristics for a one-channel graphene device (graphite walls and 2L graphene spacers). Inset: AFM image of graphene spacers in such a device (disconnected nano-cavities are seen next to the ion channel). Scale bar: 1 μm. **b,** *I-V* characteristics for a similar all-graphene device but with 200 channels. The curves at low concentrations ($10^{-2}$ and $10^{-3}$ M) are multiplied by the color-coded factors for readability. At high biases, *I-V*'s show some nonlinearity. Its origin remains to be understood. **c,** Ionic conductance for these two types of devices at various KCl concentrations, normalized per channel. Dashed-dotted line: Conductance expected from bulk conductivity of the KCl solutions. Solid lines: Leakage limits for the devices (color-coded).

**S4 Exclusion of large ions**

To make sure that our fluidic devices functioned properly and did not allow any spurious currents, we also tested them using chloride solutions with large organic cations. To this end, we chose tetramethylammonium chloride (TMACl), tetraethylammonium chloride (TEACl) and tetrabutylammonium chloride (TBACl). The conductance of TMACl was found to be ~ 50 times lower inside the Å-slits compared to that in the bulk solution (fig. S4a). TMA$^+$ has $D_H \approx 12.5$ Å and, despite this value being twice larger than $h$, the cation could still permeate through our Å-slits, albeit with rather small conductivity. On the other hand, no conductance could be detected for TEACl and TBACl solutions within our detection limits (fig. S4b). This behavior can be attributed to high activation barriers for dehydration of the large cations, which become too high to allow any discernable permeation. Note that it is unclear whether the invoked model of squashed hydration shells remains valid for, e.g., TBA$^+$ ($D_H \approx 16$ Å) and whether the latter cation exhibits even exponentially small conductivity. Indeed, TBA's diameter is expected to remain ~ 10 Å after complete dehydration[32]. This is larger than $h$ and further squashing of the cation's hydration shell may no longer be possible.

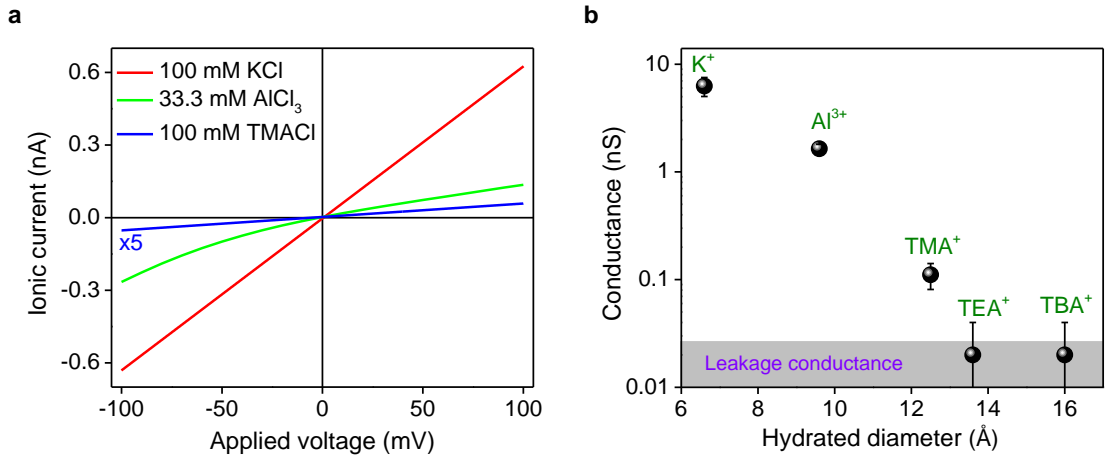

**Figure S4| Nanometer-size ions cannot permeate through Å-size slits. a**, *I-V* characteristics for a typical device (200 channels; graphite wall and 2L graphene spacer; $L$ = 3 µm) using KCl, AlCl$_3$ and TMACl solutions. **b,** The device's conductance for different chlorides. TEACl and TBACl exhibit no discernable conductance above the leakage level indicated by the shaded area.

**S5 Concentration dependence of ionic conductance**

Fig. 1d of the main text shows theoretical fits using the charge regulation model and only the ionic conductance for Å-slits with graphene walls was fitted using the constant surface charge model (Fig. 1c). For completeness, below we show fits for devices with hBN and MoS$_2$ walls using the latter model. It describes the total conductance $G$ of a fluidic channel as the sum of a bulk contribution and the one coming from surface charges as

$$G \approx \frac{FwhC}{L}\left\{\mu^+\left[\frac{\Theta}{CFh} + \sqrt{\left(\frac{\Theta}{CFh}\right)^2 + 1}\right] + \mu^-\left[-\frac{\Theta}{CFh} + \sqrt{\left(\frac{\Theta}{CFh}\right)^2 + 1}\right]\right\} \quad (S1)$$

where $w$, $h$ and $L$ are the width, height and length of channels, respectively; $F$ is the Faraday constant, $\mu^{\pm}$ the mobility of cations and anions, $C$ the salt concentration, and $\Theta$ the surface charge density. For the fits in Fig. 1c, we used $\mu^{+} = 7.08 \times 10^{-8}$ and $\mu^{-} = 2.48 \times 10^{-8}$ m$^2$ V$^{-1}$ s$^{-1}$ for K$^+$ and Cl$^-$ (see Fig. 3c), respectively. Figure S5 shows similar fits using eq. (S1) but for the case of Å-slits with hBN and MoS$_2$ walls. The theory describes our data equally well for all channel materials, independently of their chemistry. Note that we also estimated the contribution due to electro-osmosis. In our case, this conductance correction was negligibly small and could be ignored.

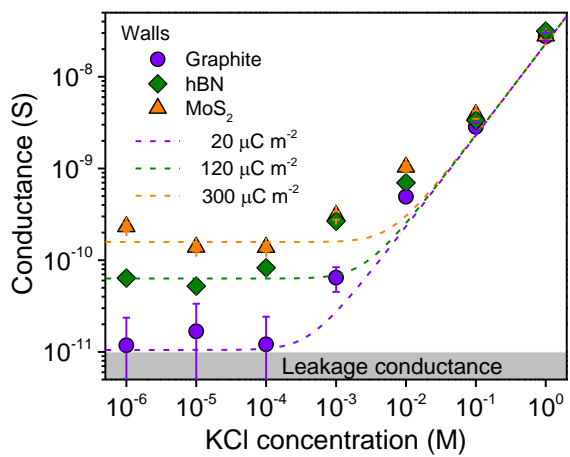

**Figure S5| Constant surface charge model for different wall materials.** The curves are fits to eq. (S1) for the data in Fig. 1d of the main text.

## S6 Effect of pH on channel conductance

To gain information about possible adsorption of OH$^-$ on capillary walls, we measured conductance of our Å-slits with hBN and graphite walls using KCl solutions and varying their pH from 2 to 10. Representative *I-V* characteristics are shown in fig. S6a. For acidic pH ≤ 6, hBN slits exhibited a nearly constant KCl conductance, which ruled out adsorption of H$^+$ on hBN surfaces. On the other hand, $G$ significantly increased for basic pH (fig. S6b). This behavior is consistent with OH$^-$ adsorption, which in turn leads to higher K$^+$ concentrations inside the slits and, hence, to the increased conductance[9]. Note that the addition of K$^+$ (from KOH) or Cl$^-$ (from HCl) to achieve the required pH also resulted in a slight (5-10%) increase in conductivity of the bulk solutions. This effect was corrected in the data presented in fig. S6. The charge density of adsorbed OH$^-$ on hBN walls was ~ 10 mC m$^{-2}$ for our highest pH = 10, which is still much lower than intrinsic charge densities $\Theta$ typical for hBN nanotubes. This observation further supports our conclusion about little intrinsic surface charge in our Å-slits (Fig. 1d).

Similar measurements with variable pH were carried out for Å-slits with graphite walls (fig. S6b). Unfortunately, graphite devices were found to quickly delaminate under basic conditions (pH > 8) which limited the experimental pH range in fig. S6b. Neither OH$^-$ nor H$^+$ adsorption could be detected on graphitic walls at the employed low salt concentrations. Note that, although our Å-slits with hBN surfaces showed better stability, large concentration gradients $\Delta \geq 100$ also resulted in their delamination. This is the reason why we restricted our diffusion experiments to $\Delta = 10$.

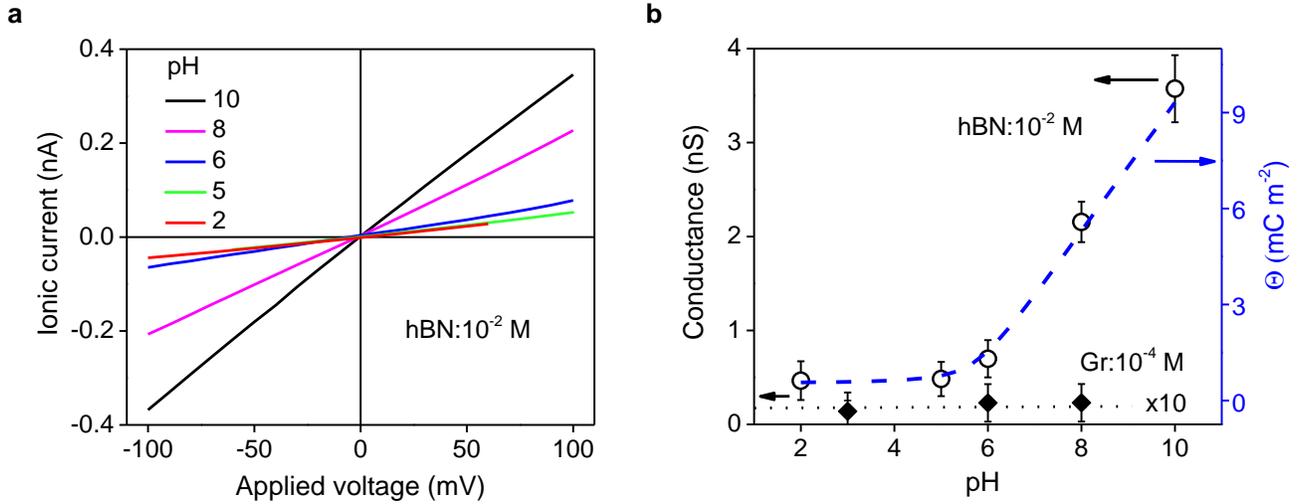

**Figure S6| pH dependent ionic conductance. a,** *I-V* characteristics for a $10^{-2}$ M KCl solution inside hBN slits (200 channels; 2L graphene spacers; $L \approx 7$ μm) at various pH. **b,** Corresponding conductance of the device (open circles). Solid diamonds: Same measurements for Å-slits with graphite walls ($10^{-4}$ M KCl). Right axis: Surface charge density Θ on hBN walls as estimated from the measured conductance. Blue dashed curve: Guide to the eye.

**S7 Control drift-diffusion experiments**

To further validate the accuracy of our analysis used to determine mobility ratios $\mu^+/\mu^-$, we studied ion transport through relatively large channels having the size $\gg D_H$ where no steric effects could be expected. The reference channels were either our silicon nitride membranes with the 3×25 μm² apertures (that is, without a tri-crystal assembly on top) or commercial PTFE membranes with multiple channels of 200 nm in diameter (*Whatman*). Standard electrodes (*HANA Instrument*) were used to avoid redox potentials. For the concentration gradient $\Delta = 10$, we measured the zero-current potential $E_m$ of, for example, about +0.5 mV for KCl, -16 mV for LiCl, -19 mV for CaCl$_2$ and -21 mV for AlCl$_3$ for both reference channels. The found $E_m$ can be described in terms of the liquid-junction potential arising from different mobilities of cations and anions in bulk solutions[25,33]. The mobility ratios $\mu^+/\mu^-$ were obtained using the Henderson formula (eq. (1) of the main text), and the extracted values agree well with those reported in the literature[24] for bulk chloride solutions (fig. S7).

**S8 Henderson vs GHK analysis using reference channels**

In the analysis given above and in the main text, we employed the Henderson equation to find $\mu^+/\mu^-$. In the literature, the alternative Goldman-Hodgkin-Katz (GHK) description[34] is also used frequently to analyze zero-current potentials arising in ion transport through nanochannels (mostly for KCl solutions). The GHK model[34] gives

$$\frac{\mu^+}{\mu^-} = -\left( \frac{\Delta - e^{z_- \frac{FE_m}{RT}}}{\Delta - e^{z_+ \frac{FE_m}{RT}}} \right) \left( \frac{1 - e^{z_+ \frac{FE_m}{RT}}}{1 - e^{z_- \frac{FE_m}{RT}}} \right) \quad \text{(S2)}$$

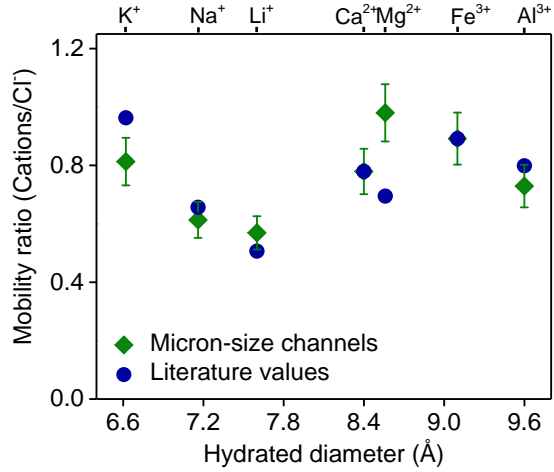

**Figure S7 | Bulk ion mobilities.** Comparison of the mobility ratios found for reference channels with the literature values.

which is an expression different from the Henderson equation (eq. (1) of the main text). We used both models to analyze our results for Å-slits and found surprisingly close values of $\mu^+/\mu^-$ for the employed concentration gradient $\Delta$ of 10 (see section S9). Nonetheless, to determine which of the two models is best for describing our experiments, we carried out additional drift-diffusion measurements, using our reference (non-confining) channels over a wide range of $\Delta$. The results are shown in fig. S8. One can see that the measured $E_m$ depended strongly on both $\Delta$ and ion species. To describe the results, we first use the Henderson expression[25]

$$E_m \approx \left( \frac{z_+/z_- + \mu^+/\mu^-}{1 + \mu^+/\mu^-} \right) \frac{1}{z_+} \frac{RT}{F} \ln \Delta \tag{S3}$$

Note that this formula leads to eq. (1) of the main text for the mobility ratio $\mu^+/\mu^-$.

As for GHK analysis, it is straightforward to solve eq. (S2) for the case of monovalent ions and obtain $E_m$ as a function of $\Delta$ and $\mu^+/\mu^-$. This yields the analytical expression

$$E_m = -\frac{RT}{F} \ln \left[ \frac{\Delta + \mu^+/\mu^-}{\Delta (\mu^+/\mu^-) + 1} \right] \tag{S4}$$

For bi- and tri- valent ions, we solved the corresponding quadratic and cubic equations numerically to obtain the dependences of $E_m$ on $\Delta$. The results for the GHK and Henderson models are compared in fig. S8, using the known bulk values of $\mu^+/\mu^-$. It is clear that the Henderson equation provides notably better agreement than the GHK model, and the discrepancy between the latter model and the experiment becomes most pronounced for large $\Delta$ and trivalent ions. This is the reason why we used the Henderson analysis in the main text. We believe that the origin of this discrepancy lies in the fact that the GHK theory does not consider additional concentration gradients along channels, which appear because of different mobilities for cations and anions, whereas the Henderson theory treats concentration gradients for anions and cations explicitly.

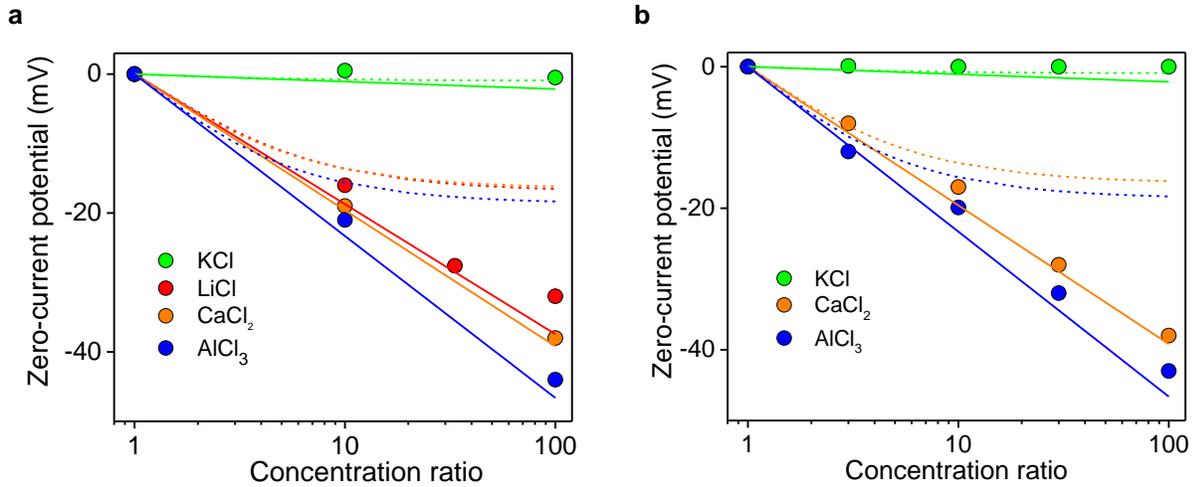

**Figure S8| Comparison between Henderson and GHK models using reference channels**. Symbols: Experimental $E_m$ for (**a**) µm-size silicon nitride apertures and (**b**) 200-nm PTFE channels. The dependences given by the Henderson and GHK models are shown by the color-coded solid and dashed curves, respectively.

### S9 Henderson vs GHK analysis for Å-slits

One can see from fig. S8 that both Henderson and GHK models provide close agreement with the experiment if modest concentration gradients $\Delta$ were used for the reference (non-confining) channels. Therefore, despite the discussed discrepancy at high $\Delta$, it is still instructive to compare the two analyses for the case of our Å-slit devices and $\Delta = 10$. We find that both models yield qualitatively similar dependences of $\mu^+/\mu^-$ as a function of $D_H$, and the difference between them reaches only a factor of 2 for trivalent salts (fig. S9).

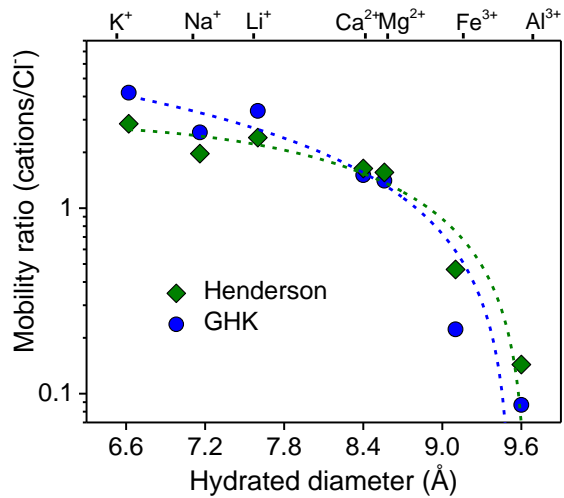

**Figure S9| Mobilities found using Henderson and GHK formulas for Å-confined ion transport.** The particular example is for a device with graphite walls and a 1L MoS$_2$ spacer. Dashed curves: Guides to the eye. Note that we were unable to determine $\mu^+/\mu^-$ for TMACl solutions because their very low conductance inside the Å-size channels disallowed accurate measurements of $E_m$.

**S10 Cation-anion asymmetry in Å-size channels**

To qualitatively illustrate the suggested model for the observed difference in mobility of confined $K^+$ or $Cl^-$ ions (by a factor of 3; see the main text), we refer to the schematic of fig. S10. The two ions have the same hydrated diameter and practically same mobilities in bulk solutions but water molecules in their hydration shells are polarized in such a way that $H^+$ and $OH^-$ groups point outside (in our case, towards the channel walls) for $K^+$ and $Cl^-$, respectively[35]. It is known that $OH^-$ groups interact with hydrophobic walls much stronger than $H^+$ (refs. 28 & 29), which can be expected to result in stronger influence of the walls on diffusion of $Cl^-$. Unfortunately, classical molecular dynamics simulations cannot describe such subtle details for ion interactions with graphene and hBN walls, and better modelling tools (e.g., quantum Monte-Carlo) need to be exploited to explain our observations quantitatively.

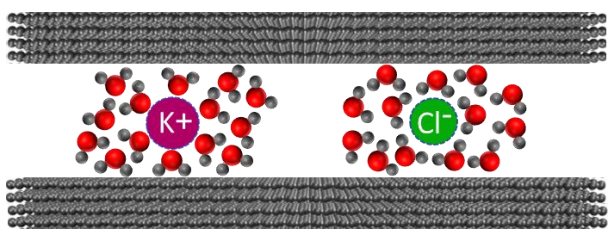

**Figure S10| Schematic of water molecules around cations and anions confined in narrow slits.** Red balls indicate oxygen, whereas hydrogen is shown in grey.

**Full list of references**